\documentstyle[prl,aps,twocolumn,psfig]{revtex}

\begin{document}

\twocolumn[\hsize\textwidth\columnwidth\hsize\csname
@twocolumnfalse\endcsname

\tighten
\draft
\title{
TeV Neutrinos from Companion Stars of Rapid-Rotating Neutron Stars
}

\author{Shigehiro Nagataki}
\address{
Department of Physics, School of Science, the University
of Tokyo, 7-3-1 Hongo, Bunkyoku, Tokyo 113-0033, Japan
}


\maketitle

\begin{abstract}
We have estimated the number flux of of $\nu_{\mu}$'s which
are produced due to the hadronic interactions between the cosmic rays
coming from a neutron star and the matter in a companion star.
The event rate at 1 km$^2$ detectors of high-energy neutrinos
such as ICECUBE, ANTARES and NESTOR is also estimated to be
$2.7 \times 10^4$ events yr$^{-1}$ when the source is located at 10 kpc
away from the Earth. We have estimated the number of such a system
and concluded that there will be several candidates in our galaxy.
Taking these results into consideration, this scenario
is promising and will be confirmed by the future observations.
\end{abstract}

\pacs{ 95.85.Ry, 98.70.Sa, 95.85.Sz, 97.80.Hn
\hspace{1.6cm} UTAP-409/02, hep-ph/0202243}

]

Neutron stars have been suggested as possible sources of very high
energy cosmic rays, including the ultra-high-energy cosmic rays whose
energies are greater than $10^{20}$eV (see Refs.~\cite{blasi00} and
references therein). It is generally considered that a magnetosphere
is formed around a neutron star where the number density has the
Goldreich-Julian value~\cite{goldreich69}. It is also pointed out
that the conditions which lead to the conversion of most Poynting flux
into kinetic energy flux in cold, relativistic hydromagnetic winds
exist around the magnetosphere (i.e. the wind zone)~\cite{begelman94}.

On the other hand, it is considered that most of the normal stars are
born in binaries. Observationally, it is reported that somewhere
between 50 and 100$\%$ of the normal stars are in binaries~\cite{abt83}.
Thus it is very natural to consider the situation that a binary is
composed of a neutron star which emits high energy cosmic rays and a
red super giant whose radius is $\sim 10^{14}$ cm. What happens if the high
energy cosmic rays rush into the companion? This is the theme in this
paper. 
It is noted that the wind from the supergiant is much weaker
than the pulsar wind and should be blown off. Thus it is not necessary to
consider the effects of interaction between the wind from the companion
star and the pulsar wind.

The conclusion is as follows. Charged pions are produced due to the
hadronic interactions between the cosmic rays and the stellar matter.
Before losing their energies, the pions can decay into leptons, including
high energy (TeV) $\nu_{\mu}$'s. The $\nu_{\mu}$'s can be detected
by 1 km$^2$ detectors of high-energy neutrinos such as ICECUBE, ANTARES
and NESTOR~\cite{halzen99}. The expected event rate is $2.7 \times 10^4$
events yr$^{-1}$.
Also, we have estimated the number of such a system
and concluded that there will be several candidates in our galaxy.
It is noted that high energy neutrinos from Cygnus X-3 were
calculated on the basis of the binary model in some excellent
papers~\cite{kolb85},
but the companion star was assumed to be a main-sequence star whose
column density seems to be too high to produce high energy neutrinos
(see Eq.~\ref{eq:18}). In this study, the companion star is assumed
to be a red super giant so as to solve the difficulty and neutrino flux
is estimated using the result of the theoretical accerelation model of
cosmic rays around the magnetosphere~\cite{blasi00},~\cite{beall01}.
Also, we estimate the number of candidates in our galaxy.

From now on, the energy spectrum of high energy cosmic rays is
reconsidered. Then the energy loss processes for the cosmic rays
and their daughters in a companion star are investigated. Next, the number
flux of $\nu_{\mu}$'s at the Earth from a single source is estimated.
Finally, the event rate and number of candidates are discussed.


Here the energy spectrum of high energy cosmic rays is estimated
(see Ref.~\cite{blasi00},~\cite{beall01} for details).
The Goldreich-Julian value can be expressed as
$
    {n_{\rm GJ}(r)} =
    B(r) \Omega /(4 \pi Z e c),
$
where $B(r)$ is the strength of the magnetic field, $\Omega$ is the
angular velocity of the neutron star. It is assumed that the magnetic
field in the magnetosphere is dominated by the dipole component and
can be expressed as 
$
    {B(r)} = B_s (R_s/r)^3,
$
where $B_s$ and $R_s$ are the strength of the surface magnetic field
and radius of the neutron star, respectively. In this study, we set
$R_s = 10$ km throughout.
The light cylinder radius can be expressed as
$
    {R_{\rm lc}} =
    10^7 \Omega_{3k}^{-1} \rm \;\; cm,
$
where $\Omega_{3k}$ is $\Omega/3000$ rad s$^{-1}$.
The typical energy of the accelerated cosmic rays $E_{\rm cr}$ can be
estimated by assuming that the equipartition of energy is realized
between the magnetic fields and cosmic rays ($E_{\rm cr} = B(r)^2/8
\pi n_{\rm GJ}(r)$). At $R_{\rm lc}$ in the equatorial plane,
$E_{\rm cr}$ can be written as
\begin{eqnarray}
    {E_{\rm cr, 0}} =
    1.5 \times 10^{9} Z B_{s,12} \Omega_{3k}^2 \rm \;\; GeV,
    \label{eq:4}
\end{eqnarray}
where $B_{s,12}$ is $B_s/10^{12}$ Gauss.
The strength magnetic field at $(R_{\rm lc},h)$ in the cylindrical
coordinate, where $h$ is the height measured from the equatorial
plane, can be expressed as
$
    {B(R_{\rm lc},h)} \sim
    B((R_{\rm lc},0)) \cos \alpha,
$
where $\cos \alpha = R_{\rm lc} / (R_{\rm lc}^2 + h^2)^{1/2}$.
Thus $E_{\rm cr}$ depends on $h$ and it can be expressed as
$
    {E_{\rm cr}(h)} \sim
     E_{\rm cr, 0} \cos^3 \alpha.
$
Thus the energy spectrum of high energy cosmic rays from 
the light cylinder can be expressed as 
\begin{eqnarray}
     \frac{dN}{dE_{z}dt}  =
      \frac{1.1 \times 10^{35} B_{s,12} \Omega_{3k}^{2}}{Z
     (E_{\rm cr, 0} E_z^2 )^{1/3}  \left[(E_{\rm cr, 0} / E_z)^{2/3}
     -1  \right]^{1/2}  },
    \label{eq:7}
\end{eqnarray}
where $E_z$ represents the energy of the cosmic rays whose charge
number is $Z$. In this study, we consider two cases for the chemical
composition of the cosmic rays, that is, protons and irons.
The chemical composition of the cosmic rays may be dominated by protons,
which will be ejected from the companion star and the magnetosphere
may be filled with the protons. On the other hand, the chemical composition
of the cosmic rays may be dominated by irons, which will be ejected from
the surface of the neutron star.


Since the energy spectrum of cosmic rays can be obtained, we consider
the energy loss processes for the cosmic rays and their daughters in a
companion star. At first, we assume that the chemical composition of
the cosmic rays is dominated by protons. We consider the energy losses
due to ionization, synchrotron radiation, pair production due to
proton-electron interaction,
photopion production, proton-proton interaction, respectively.
In order to calculate the energy loss timescale, the conditions in
the companion star has to be specified. Here we assume that the
radius and temperature of the red super giant is $10^{14}$cm and 3000K,
respectively. Thus the number of photons in the star can be estimated
as
\begin{eqnarray}
     n_{\gamma} 
                &=& 5.4 \times 10^{11} \left( \frac{T}{3000 \rm K}  
        \right)^3 \;\;
              \rm cm^{-3}.
    \label{eq:9}
\end{eqnarray}

The average baryon number density is used in this study for simplicity
and is estimated as
\begin{eqnarray}
     n_B &=& 2.8 \times 10^{15} \left( \frac{M}{10 M_{\odot}} \right)
                  \left( \frac{10^{14} {\rm cm}}{R_{\rm star}} \right)^3 \;\;
                  \rm \; cm^{-3},
    \label{eq:10.2}
\end{eqnarray}
where $R_{\rm star}$ is the radius of the companion, $M$ is the
mass of the companion and $M_{\odot}$ is the solar mass.

Now we investigate the energy loss timescale.
As for the energy loss rate due to coulomb interactions, collisions are
dominated by the thermal electrons~\cite{lang78}. Thus the energy
loss rate can be written approximately as
\begin{eqnarray}
     - \left( \frac{dE}{dt}   \right) \sim \frac{30 c \sigma_{\rm T} 
     m_e c^2 Z^2}{\beta} n_e
      \;\; \rm eV \; s^{-1}, 
    \label{eq:10.3}
\end{eqnarray}
where $\sigma_{\rm T}$ is the Thomson cross section, $\beta$ is the
velocity of the cosmic ray divided by the light speed.
Thus the energy loss timescale can be estimated as
\begin{eqnarray}
     t_{\rm ion} 
                 &\sim& 3.2 \times 10^6 \frac{1}{Z^2} \left(  
                 \frac{E}{10^{15} {\rm eV}  }  \right)
                 \left(   \frac{10^{15} {\rm cm^{-3}}  }{n_e} \right) 
                 \; \rm s  
    \label{eq:10.6}
\end{eqnarray}
which is, as we show below, too large to be effective.

As for the synchrotron energy loss timescale, it can be written as
\begin{eqnarray}
	t_{p, \rm syn} &=& \left( \frac{m_{p}}{m_e} \right)^4 \times
                         t_{e, \rm syn} \sim
 1.1 \times 10^{13} t_{e, \rm syn}, 
    \label{eq:16.2}
\end{eqnarray}
where $t_{e, \rm syn}$ is the synchrotron energy loss timescale for
an electron and can be written as
$
	t_{e, \rm syn} = 3.9 \times 10 \left( 1 \; {\rm GeV} / E
          \right) \left(  10^2 \; {\rm G}/ B  \right)^2 \; \rm s.
$
Thus synchrotron energy loss is negligible.

As for the pair production process $p$ + $\gamma$ $\rightarrow$ $p$ +
$e^-$ + $e^+$, the energy loss timescale can be written as
\begin{eqnarray}
	t_{\rm pair} 
                     &=& \frac{\pi}{16 c}\frac{m_p}{m_e}
        \frac{1}{\alpha r_e^2} \left(  \frac{c \hbar }{kT}  \right)^3
        \left(  \frac{m_e c^2}{\Gamma kT}  \right)^2
        \exp \left(  \frac{m_e c^2}{\Gamma kT}  \right) \\ 
	    &\sim& 3.1 \times 10^{19} 
        \left(   \frac{1 {\rm GeV}}{E_p}  \right)^2 
	\exp \left[ 10^6 
            \left(  \frac{2 {\rm GeV} }{E_p} 
            \right)    
            \right] \rm s ,
    \label{eq:16.4}
\end{eqnarray}
where $\alpha$ is the fine-structure constant, $r_e$ is the classical electron
radius, $\Gamma$ is the lorentz factor of proton and temperature is assumed
to be 3000K. This timescale is too large to be effective in the energy
range where we are interested in.


As for the photopion process, the $\Delta$-resonance occurs at~\cite{waxman00}
\begin{eqnarray}
	\epsilon_p = 2 \times 10^8 \left( \frac{1 \; {\rm eV}}
        {\epsilon_{\gamma}} \right) \;\; \rm GeV.
    \label{eq:13}
\end{eqnarray}
At the $\Delta$-resonance, the cross section becomes about 5$\times10^{-28}$
cm$^{-2}$~\cite{pdg02}. Even if this value is adopted, the interaction
length for this process can be estimated as
\begin{eqnarray}
	l_{p \gamma} = \frac{1}{5 \times 10^{-28} n_{\gamma} } 
                     \sim 3.7 \times 10^{15} \left( \frac{3000{\rm K}}{T}
                      \right)^3 \;\; \rm cm. 
    \label{eq:14}
\end{eqnarray}   
Thus energy loss timescale due to photopion production is
\begin{eqnarray}
	t_{p \gamma} \sim l_{p \gamma}/c = 1.2 \times 10^{5}
         \left( \frac{3000{\rm K}}{T} \right)^3 \;\; \rm s. 
    \label{eq:15}
\end{eqnarray}

On the other hand, the cross section of pion production due to
proton-proton interaction is about 50mb~\cite{pdg02}. Thus the
energy loss timescale due to proton-proton interaction is
\begin{eqnarray}
	t_{p p} \sim \frac{1}{\sigma_{pp} n_B c}
                 \sim 2.4 \times 10^{-1}
                \left( \frac{10 M_{\odot}}{M} \right)
         \left( \frac{R_{\rm star}}{10^{14} {\rm cm}} \right)^3 \; \rm s.
    \label{eq:16}
\end{eqnarray}
This timescale is much shorter than the other timescales.
Also, the interaction length $l_{p p} = c t_{p p}$ is much shorter
than the radius of the companion.

From the above discussion, we conclude that the dominant energy loss process
is proton-proton interaction and charged pions are produced due to the
interactions. The high energy protons lose $\sim 50 \%$ of their energies
per each interaction.
Here we have to investigate whether these pions can decay
into leptons before losing their energy. If they can, the leptons also
energetic and the energy of each neutrino is about 25$\%$ of the
pion's energy. Since the lifetime of a charged pion
in its rest frame is 2.6$\times 10^{-8}$ s, the observed lifetime is
\begin{eqnarray}
	t_{\pi^+} = 1.8 \times 10^{-1} \left( \frac{E}{10^{15} \; {\rm eV}} 
\right) \; \rm s.
    \label{eq:17}
\end{eqnarray}
On the other hand, the cross section of proton-pion interaction
is about 30mb~\cite{pdg02}. Thus the energy loss timescale due to
this interaction can be estimated as
\begin{eqnarray}
	t_{p \pi^+} \sim \frac{1}{\sigma_{p \pi} n_B c}
                 \sim 4.0 \times 10^{-1}
                \left( \frac{10 M_{\odot}}{M} \right)
     \left( \frac{R_{\rm star}}{10^{14} {\rm cm}} \right)^3 \; \rm s.
    \label{eq:18}
\end{eqnarray}
Thus, charged pions with energy below $\sim 2 \times 10^{15}$ eV can decay
into leptons before losing their energy.

We also check the opacity of the companion star for the high energy
neutrinos. The neutrino-nucleon total cross section
can be written approximately as
$
	\sigma_{\nu, N} \sim 5 \times 10^{-39} \left(  
       E_{\nu} / {\rm 1\; {\rm GeV}}  \right)
 \; \rm cm^2
$
~\cite{quigg86}.
Thus the average number of interaction for a single neutrino in the companion
star can be estimated as
\begin{eqnarray}
	l_{\nu, N} = 1.4 \times 10^{-9} \left(  \frac{M}{10 M_{\odot}} \right)
        \left(  \frac{10^{14} {\rm cm}  }{R_{\rm star}}    \right)^2
        \left(  \frac{E_{\nu}}{1 {\rm GeV}}   \right).
    \label{eq:18.3}
\end{eqnarray}
From this estimate, we can regard that the neutrinos can escape
from the companion freely.


Now we can estimate the number flux of
$\nu_{\mu}$'s [s$^{-1}$ cm$^{-2}$ GeV$^{-1}$]
at the Earth from a single source. It is assumed that the source is
located at 10 kpc away from the Earth ($D$ = 10kpc). 
The expected number flux is ($E_{\nu} \le 5 \times 10^{14} \rm \; eV$)
\begin{eqnarray}
    &f(E_{\nu})& = \frac{1}{4 \pi D^2} \left. \frac{dN_{\nu}}{dE_{\nu}dt}
    \right|_{E_{\nu}}  = 
    \frac{ N^2(E_p)}{4 \pi D^2}\left. \frac{12dN_p}{dt d E_{p}}
    \right |_{E_p} \\ 
    &=& 8.5 \times 10^{-17} \left( \frac{10 \rm{kpc}}{D}
        \right)^2 B_{12}^{1/3} \Omega_{3k}^{2/3} Z^{-5/3} 
        \frac{N^2(E_p)}{  E_{p}^{1/3}},
    \label{eq:19}
\end{eqnarray}
where $N(E_p) = 0.61 + 0.56 \ln (1.88E_p) + 0.129 \ln^2(1.88 E_p)$ is
mean value of the charged-particle multiplicity as a function of
$\sqrt{s}$~\cite{breakstone84}. It is noted that the mean energy of
$\nu_{\mu}$ can be written as $E_p/12N(E_p)$ when the average number of
$\pi^{\circ}$, $\pi^{+}$ and $\pi^{-}$ produced per one inelastic
interaction is same with each other.
As discussed above, there may be a cut-off around $E_{\nu} \sim 5 \times
10^{14}$ eV
due to the interaction of the charged pions and stellar matter. Of course,
such a cut-off energy strongly depends on the radius of the companion
(see Eq.~\ref{eq:18}). 
The neutral pions produced by $pp$ collision decay into gamma-rays,
which may also produce high energy pions through the $p \gamma$ interactions.
Thus the present estimation for the flux of neutrinos will be a lower limit.
Detailed Monte Carlo calculation is needed to obtain the precise flux
of neutrinos,
which is now underway and is presented in the forthcoming paper.
It is also noted that the high energy gamma-rays can not penetrate
the companion star and lose their energy by interactions with the matter
in the companion.

Finally, the event rate for a single source at 1 km$^2$ detectors
of high-energy neutrinos is estimated. The probability that a muon neutrino
will produce a high-energy upward moving muon with energy above 2 GeV in
a terrestrial detector is
approximately given by the ratio of the muon range to the neutrino mean
free path~\cite{gaisser95}. The result can be fitted as
$
	P(E_{\nu}) = 1.3 \times 10^{-9} \left(E_{\nu} /1 \; {\rm TeV}
                     \right)^{\beta}
$
where $\beta =2$ for $E_{\nu} \le 1$ TeV and $\beta =1$ for $E_{\nu} \ge 1$
TeV. 
Thus the expected event rate R [yr$^{-1}$] is (over 2$\pi$ sr)
\begin{eqnarray}
  R &=& 3\times10^{17}\left( \frac{d}{1 \; \rm {km}} \right)^2 
     \int_{2 \rm{GeV}}
       d E_{\nu} 2 \frac{2\pi}{4 \pi}
       f(E_{\nu}) P(E_{\nu}) \\
  &=& 2.7 \times 10^4 \left( \frac{d}{1 \; \rm {km}} \right)^2
      B_{12}^{1/3} \Omega_{3k}^{2/3} Z^{-5/3}
 \; \rm events \;\; yr^{-1},
    \label{eq:21}
\end{eqnarray}
where $d$ is the diameter of the detector. The factor 2 in the integral
is introduced to count in both of contributions from $\nu_{\mu}$'s and
$\bar{\nu}_{\mu}$'s. Strictly speaking, the factor (5.6/4$\pi \sim 0.45$; see
Eq.~(\ref{eq:27})) which takes into account the solid angle measured
from the companion star should be multiplied in Eq.~(\ref{eq:21}).
Since the number of detected atmospheric neutrino
background by the planned neutrino telescopes whose angular resolution $\theta$
is about $1^{\circ}$ is estimated as
$N_{\rm bkg} \sim 10 \left(\theta/ 1 {\rm deg}   \right)^2
        \; \rm km^{-2} \; yr^{-1}$~\cite{livinson01},
the signal from the companion star will be easily detected.

We add a comment on the chemical composition of the high energy cosmic
rays. In the above discussions, the chemical composition is assumed to
be proton. In the case that the chemical composition is iron, we have to
estimate the fragmentation timescale. There is an empirical formula
for the total inelastic cross-section of protons on nuclei with mass number
$A \ge 1$~\cite{letaw83}. According to the formula, the cross section can be
written as
$
	\sigma_{\rm tot} = 45 A^{0.7} \left[  1 + 0.016 \sin (5.3 - 2.63 \ln A)    \right] \; \rm mb.
$
Thus the fragmentation timescale can be estimated as
$	\tau_{\rm f} = 3.8 \times 10^{-2} \left( 56/A \right)^{0.7}
                \left(   10^{15} \; \rm cm^{-3}/n_{\rm B}   
         \right) \;\; \rm s.
$
This timescale is shorter than the proton-proton interaction timescale
(see Eq.~\ref{eq:16}). Even if the chemical composition is composed of
deuterium ($A=2$), the fragmentation timescale is comparable with the
proton-proton interaction timescale as long as $n_{\rm B} \sim 3 
\times 10^{15}$ cm$^{-3}$. Thus we can conclude that the fragmentation
occurs at first when the cosmic rays rush into the companion star and
subsequent reactions are same with the case in which chemical composition
of the cosmic rays is proton. As for the number flux of nucleus
after the fragmentation, it will be $(A/Z)^{5/3} \sim 3.2$ times larger
than the proton
case (see Eq.~\ref{eq:7}). Thus the expected event rate of high energy
neutrinos will be also enhanced by a factor of $\sim 3.2$.

When the neutrino mixing angles are really large~\cite{fukuda98}, the 
fluxes of $\nu_{\mu}$'s and $\bar{\nu}_{\mu}$'s will fall off
by a factor of 3. In this case, $\nu_{\tau}$'s might be detected
as the double bang events~\cite{athar00}.

We also present a rough estimate for the number of candidates (N) in our
galaxy. It may be expressed in the form
\begin{eqnarray}
  N &=& \tau \times {\rm SNR} \times f \times \left( \frac{d \Omega}{4 \pi}
        \right),
    \label{eq:22}
\end{eqnarray}
where $\tau$ is the duration [yr] for a source to emit the high-energy
$\nu_{\mu}$'s, SNR is the supernova rate [yr$^{-1}$] in our galaxy
at present, $f$ is the probability that a massive star which forms a
neutron star at its death has a companion whose mass is above
$\sim 8M_{\odot}$, $d \Omega$ is the average solid angle of a companion star
measured from the neutron star.

The duration $\tau$ will be given as $\tau$ = min($\tau_{\rm star}$,
$\tau_{\rm NS}$), where $\tau_{\rm star}$ is the lifetime of the massive
companion star ($\sim 10^6$yr) and $\tau_{\rm NS}$ is the duration in which
the neutron star can emit the high energy cosmic rays. From Eq.~\ref{eq:17},
the energy of the cosmic rays should be greater than $\sim 10^{16}$ eV
in order to produce much energetic neutrinos.
On the other hand, the angular velocity of a neutron star decreases as
\begin{eqnarray}
  \Omega (t) &=& \frac{1}{ \sqrt{ \frac{1}{\Omega_i^2} + 
                 \frac{B_s^2R_s^6}{3c^3I}t }} \;\; \rm rad \; s^{-1},
    \label{eq:23}
\end{eqnarray}
where $\Omega_i$ is the initial angular velocity and
$I \sim 10^{45}$g cm$^{2}$ is the inertial moment of a neutron star.
From Eq.~\ref{eq:4}, $E_{\rm cr, 0}$ becomes $\sim 10^{16}$ eV
when $B_{s,12} = 1$ and $ \Omega_{3k} = 0.1$.
Thus, if $\Omega_i$ is much larger than 0.1, $\tau_{\rm NS}$ can be
estimated as
\begin{eqnarray}
  \tau_{\rm NS} &\sim&  3 \times 10^{4}
  \left( \frac{10^{12} \; {\rm G}  }{B_s}  \right)^2 \;\; \rm yr,
    \label{eq:24}
\end{eqnarray}
which is shorter than $\tau_{\rm star}$.
Thus it is concluded that $\tau$ = $\tau_{\rm NS}$.

The present SNR is about $10^{-2}$ yr$^{-1}$~\cite{porciani01}.
As for $f$, since most of the normal stars are considered
to be born in binaries~\cite{abt83}, we can estimate this value
only by using the salpeter's initial mass function. The result
can be written as~\cite{madau00}
\begin{eqnarray}
 f = \frac{\int_{8}^{125} dm \; m^{-2.35}}{\int_{0.5}^{125} dm \; m^{-2.35}}
   = 2.4 \times 10^{-2}.
    \label{eq:24.2}
\end{eqnarray}

Finally, the average solid angle of a companion star measured from the
neutron star is estimated. Although there is much uncertainty,
Popova et al. (1982) discussed that the distribution of the
distance ($a$) between the binary stars may obey the relation as follows:
\begin{eqnarray}
	\frac{dN}{da} = \frac{A}{a} \;\; {\rm cm^{-1}} 
         \;\; \left( 10R_{\odot}  \le a \le 10^6 R_{\odot} \right), 
    \label{eq:25}
\end{eqnarray}
where $A$ = $(5 \ln 10)^{-1}$ is the normalization factor and
$R_{\odot}$ is the solar radius. Since the radius of the companion
is assumed to be $10^{14}$ cm, the opening angle $\theta$ can be
estimated as $\sin \theta = 10^{14}/a$ as long as $a$ is greater
than $10^{14}$cm. The solid angle can be estimated as
\begin{eqnarray}
d \Omega = \int_0^{2\pi} d \phi \int^{\theta}_0 \sin \theta d \theta 
         = 2 \pi \left[ 1 - \left(1 - \frac{10^{28}}{a^2} 
             \right)^{1/2}     \right].
    \label{eq:26}
\end{eqnarray}
In case $a \le 10^{14}$cm, we assume that
the solid angle is 4$\pi$. 
We can easily show by Eqs.~(\ref{eq:10.6})-(\ref{eq:16}) that the
dominant energy loss process is still proton-proton interaction
within $\tau \sim \tau_{\rm NS}$
even if the pulsar wind blows off the envelope of the companion star.
Now we can estimate the mean solid angle
of a companion measured from the neutron star. It can be calculated as
\begin{eqnarray}
\label{eq:27}
&d& \Omega = \int_{10 R_{\odot}}^{10^{14}} 4 \pi  \frac{dN}{da}da \\ 
         && + \int_{10^{14}}^{10^6 R_{\odot}} 2 \pi \left[ 
          1 - \left(1 - \frac{10^{28}}{a^2} \right)^{1/2}    
          \right]  \frac{dN}{da}da   \sim 5.6 \nonumber  
\end{eqnarray}  

At last, we can estimate the number of candidates (N) in our galaxy.
It is written as
\begin{eqnarray}
  N &\sim& 3 \times 10^4 \times 10^{-2} \times 2.4 \times 10^{-2}
    \times \frac{5.6}{4 \pi} = 3.2.
    \label{eq:28}
\end{eqnarray}

We have estimated the number flux of of $\nu_{\mu}$'s which
are produced due to the hadronic interactions between the cosmic rays
coming from a neutron star and the matter in a companion star.
The event rate at 1 km$^2$ detectors of high-energy neutrinos
such as ICECUBE, ANTARES and NESTOR is also estimated to be
$2.7 \times 10^4$ events yr$^{-1}$ when the source is located at
10 kpc away from 
the Earth. Finally, we have estimated the number of such a system
and concluded that there will be several candidates in our galaxy.
Taking these results into consideration, this scenario
is very promising and will be confirmed by the future observations.



\vspace{-0.5cm}


\end{document}